\documentclass[12pt]{article}           
\usepackage{hyperref}
\pdfoutput=1
\usepackage[pdftex]{graphicx}
\usepackage{amsmath,amssymb}
\usepackage{cancel}
\usepackage{appendix}
\newcommand{\gs}{\ensuremath{g_s}} 
\newcommand{\ls}{\ensuremath{l_s}} 

\def\p{\partial}

\newcommand{\cN}{{\mathcal{N}}}

\newcommand{\Xp}{{X^{'}}}
\newcommand{\Xd}{{\dot{X}}}

\newcommand{\be}{\begin{equation}}
\newcommand{\ee}{\end{equation}}
\newcommand{\bea}{\begin{eqnarray}}
\newcommand{\eea}{\end{eqnarray}}

\begin{document}

\begin{titlepage}

\begin{flushright}
DAMTP-2013-44\\
UTTG-22-13\\
 TCC-018-13
\end{flushright}

\begin{center} \Large \bf Holographic EPR Pairs, Wormholes and Radiation
\end{center}

\begin{center}
 Mariano Chernicoff$^{\star}$\footnote{mc790@cam.ac.uk},
Alberto G\"uijosa$^{\dagger}$\footnote{alberto@nucleares.unam.mx}
and Juan F.~Pedraza$^{\natural}$\footnote{jpedraza@physics.utexas.edu}

\vspace{0.2cm}
$^{\star}$ Department of Applied Mathematics and Theoretical Physics,\\
University of Cambridge, Cambridge CB3 0WA, UK\\
\vspace{0.2cm}
${}^{\dagger}$ Departamento de F\'{\i}sica de Altas Energ\'{\i}as, Instituto de Ciencias Nucleares, \\
Universidad Nacional Aut\'onoma de M\'exico,
\\ Apartado Postal 70-543, M\'exico D.F. 04510, M\'exico\\
\vspace{0.2cm}
$^{\natural}$ Theory Group, Department of Physics and Texas Cosmology Center,\\
University of Texas at Austin, Austin, TX, 78712, USA\\
\vspace{0.2cm}
\end{center}

\begin{center}
{\bf Abstract}
\end{center}
\noindent As evidence for the ER=EPR conjecture, it has recently been observed that the string that is holographically dual to an entangled quark-antiquark pair separating with (asymptotically) uniform acceleration has a wormhole on its worldsheet. We point out that a two-sided horizon and a wormhole actually appear for much more generic quark-antiquark trajectories, which is consistent with the fact that the members of an EPR pair need not be permanently out of causal contact. The feature that determines whether the causal structure of the string worldsheet is trivial or not turns out to be the emission of gluonic radiation by the dual quark and antiquark. In the strongly-coupled gauge theory, it is only when radiation is emitted that one obtains an unambiguous separation of the pair into entangled subsystems, and this is what is reflected on the gravity side by the existence
of the worldsheet horizon.

\vspace{0.2in}
\smallskip
\end{titlepage}

\section{Introduction} \label{introsec}

Based on various strands of information,
including \cite{israel,bklt,maldaeternal,vrbuilding,amps,ampss,braunstein},
 Maldacena and Susskind \cite{maldasuss} recently argued that configurations of distant black holes connected by a wormhole, or Einstein-Rosen (ER) bridge, should be interpreted as states where the black holes are maximally entangled.\footnote{A closely related set of beautiful observations on the important geometric role played by entanglement in quantum gravity can be found in \cite{vrcomments,vrbuilding,vrpatch,vrrindler,vrevaporating}. See also \cite{swingle1,swingle2}.}
 They speculated that this relation might extend to other entangled systems, to the point that even a single Einstein-Podolsky-Rosen (EPR) pair would be connected by a (highly quantum) wormhole. They dubbed this conjectural connection the ER=EPR relation.

 Jensen and Karch \cite{jensenkarch} then noted that, in the context of the holographic correspondence \cite{malda,gkpw} for a strongly-coupled CFT such as $\cN=4$ super-Yang-Mills, this can be illustrated very concretely by taking the EPR pair to be a color-singlet (and therefore entangled) quark-antiquark pair that separates in vacuum with uniform acceleration. This is dual to a string in anti-de Sitter (AdS) spacetime, whose endpoints first approach and then move away from one another along the hyperbolic trajectories characteristic of constant acceleration. The relevant string solution was found in \cite{xiao},\footnote{A portion of this solution (also obtained in \cite{ppz}) is identical to the one describing an \emph{isolated} quark undergoing uniform acceleration, which was later understood \cite{brownian,noline} to be a particular case of the general embeddings worked out much earlier in \cite{mikhailov}. We will return to this connection in Section \ref{accelsec}.} and describes a semicircular string which starts out with infinite size, decreases down to a minimal radius, and then grows again as the endpoints recede from one another. The crucial point is that the induced metric on the worldsheet of this string encodes a causal structure that is the exact analog of the one considered for spacetime black holes in \cite{maldasuss}: there is a double-sided horizon at a fixed radial depth in AdS, with one exterior region for each string endpoint, and a wormhole in between. It was subsequently pointed out in
 \cite{sonner}, based on \cite{semenoffzarembo} (see also \cite{gorsky}), that the uniformly accelerating solution of \cite{xiao} is the Lorentzian continuation of the instanton associated with Schwinger pair creation in an electric field, in direct analogy with the pair production of black holes discussed in \cite{maldasuss}.
 Very recently, \cite{penna} provided interesting
 evidence (and a restriction) for ER=EPR along a different
 route, by testing entropy inequalities.

 The authors of \cite{jensenkarch} emphasized the need for the quark and antiquark to be permanently out of causal contact. This restricts the discussion to the class of trajectories where the quark and antiquark asymptotically undergo uniform acceleration, in which case the double-sided horizon on the worldsheet is inherited from the spacetime Rindler-type horizons associated with the corresponding accelerated observers \cite{xiao,hkkl,brownian}.
 Knowing that the existence of EPR correlations between the quark and antiquark requires entanglement, but not permanent causal disconnection, this raises the question of what happens for other trajectories of the EPR pair. This is the issue that we address in the present work. While there exist cases where the worldsheet causal structure remains trivial, we will find that a double-sided horizon and a wormhole do appear whenever gluonic radiation is emitted, just as was discovered in \cite{dragtime} for the case of an isolated quark.\footnote{The appearance of a worldsheet horizon had been noted previously in \cite{gubserqhat,ctqhat} for the gauge theory at finite temperature.}  Our conclusions are given in Section \ref{discussionsec}.

\section{Setup} \label{setupsec}

Our analysis applies to any CFT with a holographic dual, in any number of dimensions, but for concreteness, we will phrase the discussion in terms of the familiar duality equating $SU(N_c)$ maximally supersymmetric ($\cN=4$) Yang-Mills (MSYM) on $3+1$ dimensional Minkowski space with Type IIB string theory on the AdS$_5$  Poincar\'e patch\footnote{The features specific to this particular example, such as the existence of an $S^5$ and a Ramond-Ramond five-form flux, will play no role here, and will therefore be ignored.}
\begin{equation}\label{metric}
ds^2=G_{mn}dx^m dx^n={R^2\over z^2}\left(
-dt^2+d\vec{x}^{\,2}+dz^2 \right)~.
\end{equation}
The MSYM coupling is connected to the string coupling via $g_{YM}^2=4\pi\gs$, and the radius of curvature $R$ is related to the gauge theory 't Hooft coupling $\lambda\equiv g_{YM}^2 N_c$ through
\begin{equation}\label{lambda}
 \lambda={R^4\over \ls^4}~,
\end{equation}
 where $\ls$ denotes the string length.

A quark (hypermultiplet) sector can be added to MSYM by introducing probe
 flavor (D7-)branes in the bulk of AdS \cite{kk}. An isolated quark with mass $m$ is dual to a string extending from the Poincar\'e horizon at $z\to\infty$ to the edge of the branes at radial location
 \begin{equation}\label{zm}
 z_m=\frac{\sqrt{\lambda}}{2\pi m}~.
 \end{equation}
 A $\cap$-shaped string with both of its endpoints at $z=z_m$ is dual to a quark and antiquark in the color-singlet configuration. Morally speaking, the string endpoints represent the quark and antiquark, while the body of the string represents the color `flux tube' in between, i.e., the (near and radiation) gluonic field profile sourced by the fundamental color sources.
 In the $N_c\to\infty$, $\lambda\to\infty$ limit, the string embedding is determined by extremizing the Nambu-Goto action
\begin{equation}\label{nambugoto}
S_{\mbox{\scriptsize NG}}=-{1\over 2\pi\ls^2}\int
d^2\sigma\,\sqrt{-\det{g_{ab}}}
=-{1\over 2\pi\ls^2}\int
d^2\sigma\,\sqrt{\left(\Xd\cdot\Xp\right)^2-\Xd^2\Xp^2}~,
\end{equation}
where $g_{ab}\equiv\p_a X^m\p_b X^n G_{mn}(X)$ ($a,b=0,1$) is
the induced metric on the worldsheet, and of course $\,\dot{}\equiv\p_{\sigma^{0}}\equiv\p_{\tau}$, ${}^{\prime}\equiv\p_{\sigma^1}\equiv\p_{\sigma}$. Unless stated otherwise, we will envision the string in the static gauge $\tau=t$, $\sigma=z$. For a static $q$-$\bar{q}$ pair, this setup yields an attractive Coulomb potential between the quark and antiquark \cite{maldawilson,reyee}, as expected by the conformal invariance of MSYM, while generating no long-range disturbance in the gluonic field \cite{undulate,linshuryak}, as expected for a color-singlet.

For some of the sections below it will be useful to recall here that two general classes of solutions
to the Nambu-Goto equation of motion are known for a string with a \emph{single} endpoint
on the flavor branes, which is dual to an \emph{isolated} quark (or antiquark) in MSYM.
In the form in which they were originally
constructed in \cite{mikhailov}, they take as input an \emph{arbitrary}
timelike trajectory of the quark/endpoint
 \begin{equation}\label{quarkx}
\vec{x}(t)=\vec{X}(t,z_m)~,
\end{equation}
in the infinitely massive case $z_m=0$, and they read
\begin{eqnarray}\label{mikhsolnoncovariant}
t(t_r,z)&=&t_r\pm\frac{z}{\sqrt{1-\vec{v}(t_r)^2}}~,\\
\vec{X}(t_r,z)&=&\vec{x}(t_r)\pm\frac{\vec{v}(t_r) z}{\sqrt{1-\vec{v}(t_r)^2}}~. \nonumber
\end{eqnarray}
The upper (lower) sign refers  to a purely retarded (advanced) solution, where information
 propagates from the endpoint towards the Poincar\'e horizon (vice versa).
 More specifically,
 (\ref{mikhsolnoncovariant}) shows that $t_r$ plays the role of a retarded (advanced)
 time, and the endpoint information flows along the straight lines at fixed $t_r$,
 which turn out to be null geodesics both on the string worldsheet and in spacetime.
 The generalization to a quark with finite mass, $z_m>0$,
 was given in \cite{dragtime,lorentzdirac,damping}.

\section{Asymptotically Free Motion} \label{freesec}

In this and the following two sections, we will study various kinds of evolutions for the entangled
$q$-$\bar{q}$ pair, to figure out
under what conditions there might appear a double-sided horizon on the dual string worldsheet.
Let us first focus attention on the more physical situation
 where the external forcing of the pair does not go on
for an infinite amount of time.
Consider then a string dual to a color-singlet quark and antiquark with arbitrary
(possibly forced) trajectories up to a time
$t=t_{\mbox{\scriptsize free}}$, after which both endpoints are forever
set free.\footnote{Ultimately, it will be clear
that the conclusions would be the same if the forcing is turned
off gradually rather than at a definite time, but it is cleaner
to phrase the argument this way.} {}From this moment on,
the total energy and momentum of the string, $E_{\mbox{\scriptsize tot}}$ and
$\vec{P}_{\mbox{\scriptsize tot}}$, are conserved
(to leading order in the large $N_c$ expansion).
For any given string configuration at
$t=t_{\mbox{\scriptsize free}}$,
it will be convenient to phrase our discussion
in the Lorentz frame where $\vec{P}_{\mbox{\scriptsize tot}}=0$.

The subsequent evolution is either unbound, with the quark and antiquark generically moving
an infinite distance apart, or bound, with the pair
executing forever some kind of collective motion.
If $E_{\mbox{\scriptsize tot}}<2m$, then the system
is surely bound. For $E_{\mbox{\scriptsize tot}}\ge 2m$, the pair has enough energy to dissociate,
but,
without solving the equation of motion, we cannot take this outcome for granted,
because we are in a non-Abelian (albeit non-confining) theory, and the `flux tube'
connecting the $q$ and $\bar{q}$ could in some cases sustain a discrete tower of
excitations with arbitrarily
high energy (stable at leading order in $1/N_c$) \cite{undulate}. Nevertheless, on
physical grounds it is clear that the vast majority of solutions with
$E_{\mbox{\scriptsize tot}}\ge 2m$ will be unbound.

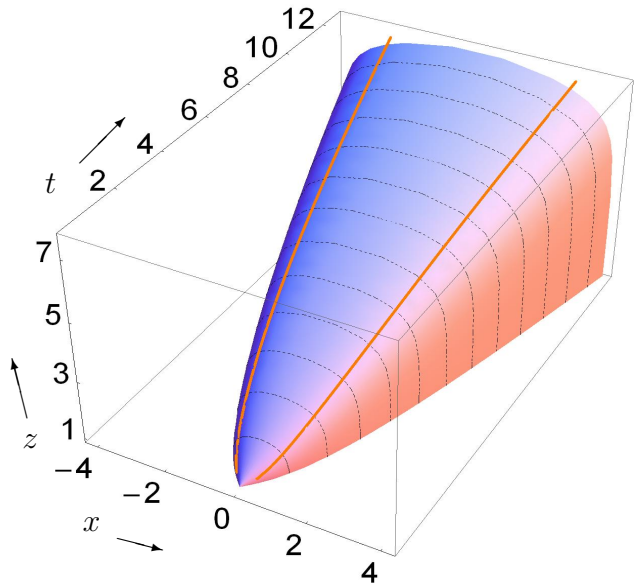
\begin{figure}[htb]
\begin{center}
\vspace*{0.5cm}
 \setlength{\unitlength}{1cm}
\includegraphics[width=8cm,height=8cm]{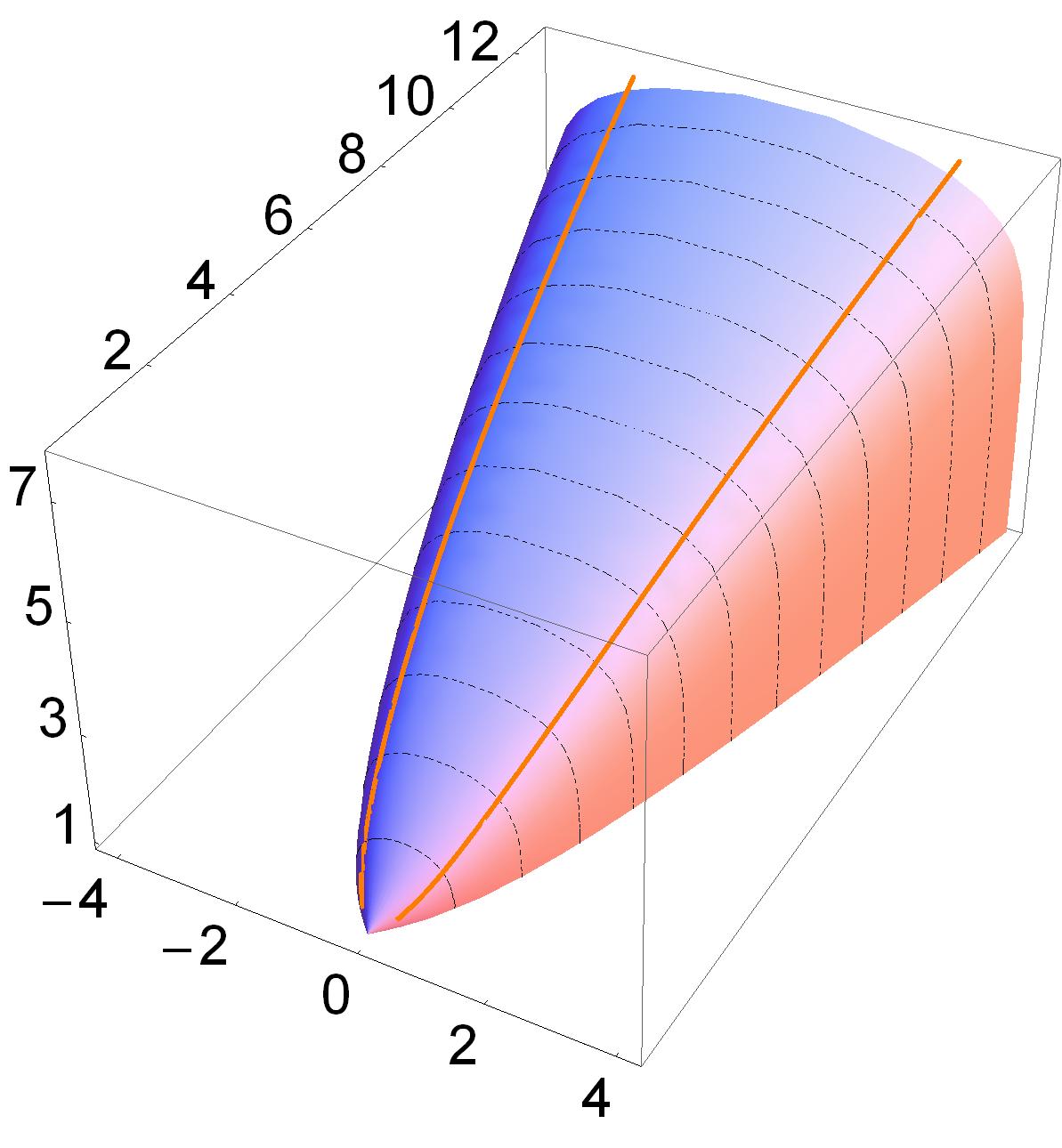}
\begin{picture}(0,0)
\put(-8.1,5.4){ $t$}
\put(-7.5,5.8){\vector(1,1){0.6}}
\put(-7.6,0.9){ $x$}
\put(-7.0,0.8){\vector(4,-1){0.6}}
\put(-8.4,2.0){ $z$}
\put(-8.2,2.4){\vector(-1,4){0.2}}
\end{picture}
\end{center}
\vspace*{-0.8cm}
 \caption{Evolution of the string dual to an initially pointlike color-singlet quark antiquark pair. The particles
start out at the origin at $t_{\mbox{\scriptsize free}}=0$, and thereafter separate back-to-back without being externally forced. The graph displays the result of numerically integrating the equations of motion for the string, with a choice of gauge and initial conditions as in \cite{hkkky,dragtime}, and a value $E_{\mbox{\scriptsize tot}}= 4m$ for the total energy of the system. The axes are given in units of $z_m$. The dotted curves depict snapshots of the string at fixed time. As explained in the main text, at late times the string is seen to approach a $\sqcap$ shape. The nearly straight solid lines (orange online) are stationary-limit curves on the worldsheet. For $t\to\infty$ these lines coincide with the worldsheet horizon (see the Appendix), and therefore serve to delimit the horizontal portion of the string which asymptotically encodes the gluonic radiation emitted by the pair.}\label{pointfig}
\end{figure}

In the unbound cases, we will end up
with a quark and antiquark moving with constant velocities, and, unless these velocities are equal,
they will be separated by an infinite distance.
Let $e_q$, $e_{\bar{q}}$ and $\vec{p}_q$, $\vec{p}_{\bar{q}}$ be the corresponding
energies and momenta. Leaving aside the marginally bound case, we will have
$E_{\mbox{\scriptsize tot}}> e_q+e_{\bar{q}}$, $\vec{p}_q\neq \vec{p}_{\bar{q}}$, meaning that
 some energy and momentum have been carried away by gluonic radiation.
On the gravity side, the duals of the free quark and antiquark will be vertical (i.e., purely radial) segments of
string translating uniformly, extending
from $z_m$ to a point arbitrarily close to the Poincar\'e horizon, so the excess energy and momentum
of the system
will necessarily be carried by the middle portion of the string,
which connects the upper ends of the vertical segments.  At asymptotically late times,
this portion will have fallen towards and almost reached the Poincar\'e horizon
(in terms of the gauge theory time, it never quite crosses, if we
ignore the
string's backreaction, which is suppressed at large $N_c$).\footnote{Were it not for this proximity
with the spacetime horizon at $z\to\infty$, the middle portion
 would not be able to extend from one vertical
segment to the other, while carrying only the (in general finite) excess energy and momentum.}
Evidently, this `horizontal' portion of the string is the embodiment of radiation, and its fall
towards $z\to\infty$ reflects, via the UV-IR connection \cite{uvir},
 the fact that the radiation travels out to infinity in the CFT.
 So, in the unbound case,
 whenever the pair emits radiation, the string will end up having essentially a $\sqcap$
 shape.\footnote{At any finite time, the transition region between the vertical and horizontal
 portions will of course be smooth.} An example is shown in Fig.~\ref{pointfig}.

 Let us now discuss the worldsheet causal structure. Readers interested in the relevant formulas are advised to consult the Appendix. At asymptotically late times, the induced metric on the vertical segments of the string
 worldsheet
 will evidently have the same causal structure as the ambient AdS spacetime.
 From any point on those segments,
 there exists an outward and an inward pointing null direction
 (respectively towards smaller or larger $z$).
 But in the horizontal portion the derivatives of the embedding fields $\vec{X}(t,z)$
 are large, and the causal structure on the worldsheet is highly distorted with respect to that in
  AdS, to the point that both null directions end up pointing toward larger $z$ above a stationary-limit curve on either side of the string (see Fig.~\ref{pointfig}).
   This is why, in general, the excess energy and momentum end up deposited there, rather than seeping
   out towards the vertical segments. So, if we start at the `corners' of the asymptotic
   $\sqcap$-shaped string and trace backwards the null geodesics that would have ordinarily pointed
   outwards,
   we will delineate a region from which no signal can escape into the two vertical segments. In
   other words, we find that at late times the worldsheet contains a double-sided event horizon.
   And, in retrospect, we also learn
  that in bound cases there can be no emission of radiation
 at leading order in $1/N_c$,
 because if a middle portion of the string falls to the
 Poincar\'e horizon, the quark and antiquark segments
 will end up causally disconnected.

 How far towards the AdS boundary (and into earlier times) the horizon extends will of course
 depend on the details of the history of the quark and antiquark prior to
 $t=t_{\mbox{\scriptsize free}}$, which we have thus far left unspecified, in the interest of
 generality. The most natural class of initial conditions
 is such that at early times there is no radiation, i.e., the system is initially bound.
 Forcing will then be required to evolve to an unbound case. {}From the worldsheet
 perspective, the interpretation is simply that, having started with no horizon, the energy we
 inject through the endpoints collapses to form a black hole, at a definite time
 $t_{\mbox{\scriptsize BH}}$. From that moment on, the quark and antiquark will be
 forever causally
 disconnected along the string,  even though, in general, they
  are not required to be permanently out of contact through spacetime.

\section{Asymptotically Uniform Acceleration} \label{accelsec}

Having seen that a double-sided horizon appears for pair histories much more general
than the case of uniform
acceleration discussed in \cite{jensenkarch},
it is instructive to revisit that particular instance. For infinitely massive
quark and antiquark with constant (magnitude of the four-)acceleration $A$,
the solution found in \cite{xiao} is
\begin{equation}\label{xiaoaccel}
X(t,z)=\pm\sqrt{A^{-2}+t^2-z^2}~,
\end{equation}
where the upper/lower sign refers to the right and left half of the string, whose endpoints we will,
for definiteness,  associate respectively with the quark and antiquark.\footnote{For
finite mass, (\ref{xiaoaccel}) is still valid after
the replacement $A^{-2}\to A^{-2}+z^2_m$ \cite{brownian}. The relation between the external
force (electric field) $F$ and $A$ is then non-trivially modified \cite{lorentzdirac,damping}, in such a way that, even
though $F$ is bounded by its critical value  (associated with pair creation), $A$ can be arbitrarily
large \cite{brownian}.}

This embedding displays a double-sided worldsheet horizon at $z=A^{-1}$, signifying
that the quark and antiquark are permanently out of causal contact along the string (as they are also through
spacetime). Since the evolution of either one is not affected by the existence of the other,
it is natural to wonder whether this embedding could fall under the class of isolated quark solutions
obtained in \cite{mikhailov}, which we recalled above in (\ref{mikhsolnoncovariant}). As it turns
out, it was shown in \cite{brownian,noline} that this is precisely the case. The key point
is that the past (future) asymptotic behavior
for uniform acceleration happens to be so violent that the string described by the
retarded (advanced) version of
 (\ref{mikhsolnoncovariant}), which ordinarily extends all the way to $z\to\infty$, as expected
 for an isolated quark, actually terminates at $z=A^{-1}$, at a point that travels at the speed of light along the line $x=-t$. The embedding then clearly needs
 to be somehow completed beyond that \cite{noline}.

\begin{figure}[tbph]
\begin{center}
\vspace*{0.5cm}
 \setlength{\unitlength}{1cm}
\includegraphics[width=7cm,height=3.5cm]{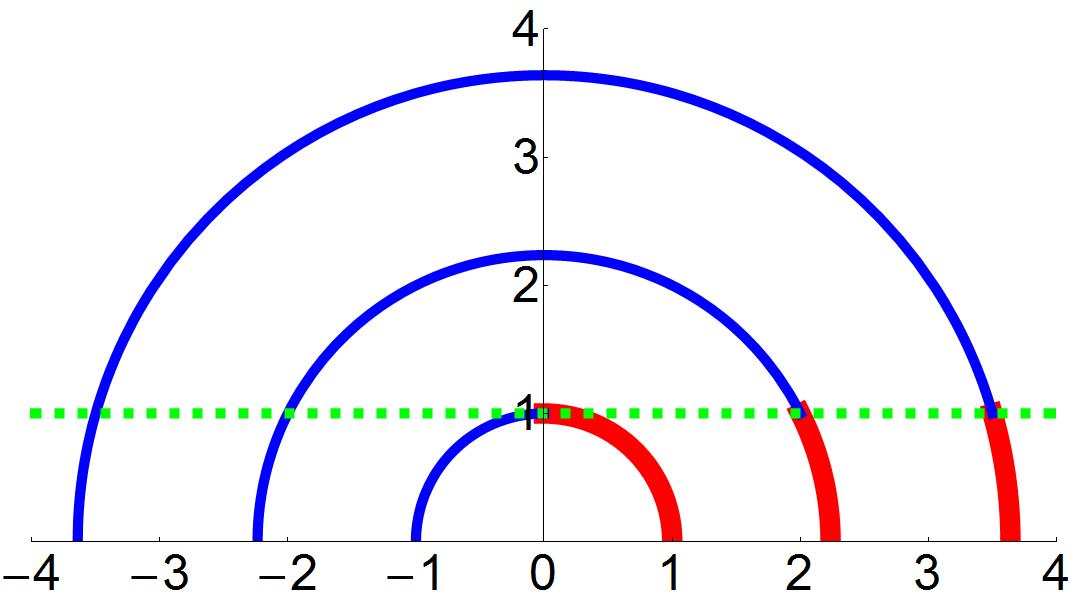}
\hspace*{0.5cm}
\includegraphics[width=7cm,height=3.5cm]{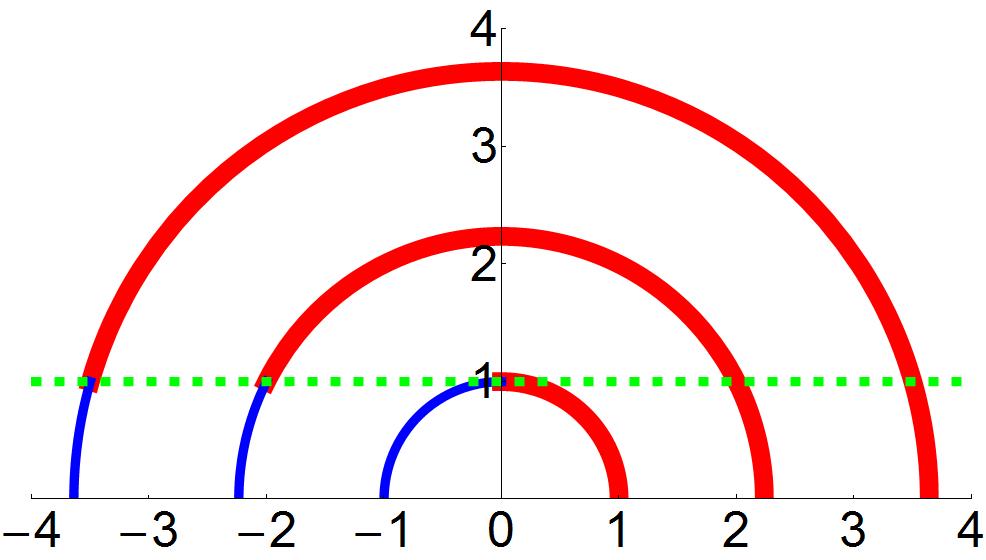}
 \begin{picture}(0,0)
   \put(0.1,0.2){$x$}
   \put(-7.7,0.2){$x$}
   \put(-11.2,3.3){$z$}
   \put(-3.4,3.3){$z$}
   \put(-13.7,1.2){\vector(-1,-3){0.15}}
   \put(-9.0,1.2){\vector(1,-3){0.15}}
   \put(-5.9,1.0){\vector(1,3){0.15}}
   \put(-1.2,1.0){\vector(-1,3){0.15}}
  \end{picture}
 \end{center}
 \vspace*{-0.5cm}
\caption{Successive snapshots of the string (\ref{xiaoaccel}) dual to a uniformly accelerated quark-antiquark pair, understood as a combination of a retarded (quark, thick red) and an advanced (antiquark, thin blue) version of the solution (\ref{mikhsolnoncovariant}), in units of $A^{-1}$. The left figure portrays $t=-3.5,-2,0$, when the semicircular string is shrinking as the endpoints slow down. The right figure depicts $t=0,2,3.5$, when the string is growing as the endpoints speed up again. Throughout, there is a double-sided horizon at $z=A^{-1}$ (dotted green), and the juncture between the retarded quark and advanced antiquark embeddings is seen to be moving uniformly to the left at the speed of light. The arrows indicate the direction of energy flow along the string.}\label{xiaofig}
\end{figure}
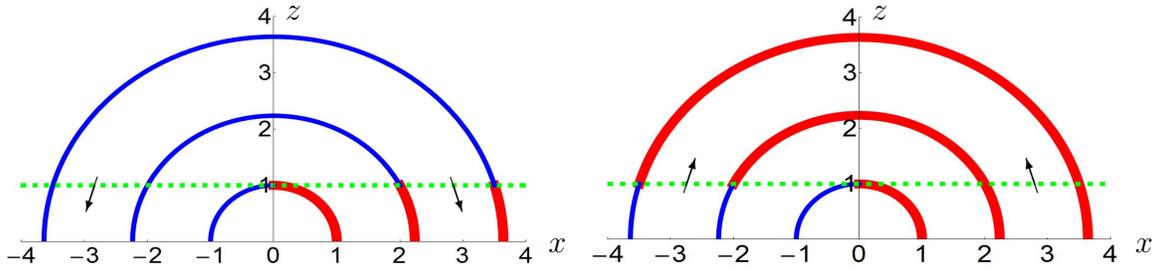

 The solution of \cite{xiao} is obtained
 upon joining together a retarded quark embedding of \cite{mikhailov}
 with an advanced antiquark embedding
 (or vice versa). A peculiar feature is that this combination is not symmetric in the quark
 and antiquark: for $t<0$, the arc corresponding to the quark (antiquark)
  covers less (more) than a quarter circle,
 and for $t>0$ the situation is reversed \cite{noline}. This is in spite of the fact that
 the net result (\ref{xiaoaccel}) is completely symmetric, as is
 the spacetime energy and momentum flow along the string (which, as expected,
 reflects energy flow out of the horizon for all $t<0$,
 when the string contains a white hole, and into
 the horizon for all $t>0$, when it is a black hole that is present).
 In this particular case, then,
 the distinction  between the retarded and advanced solutions of \cite{mikhailov} is inconsequential.
 The situation is summarized in Fig.~\ref{xiaofig}.

 Since the abrupt termination of the retarded/advanced solutions
 (\ref{mikhsolnoncovariant}) hinges only
 on the asymptotic early/late behavior of the quark/antiquark, we can easily construct
 a large family of explicit
 analytic solutions which generalize (\ref{xiaoaccel}). Given any quark and antiquark trajectories
  whatsoever
 that  asymptote to uniform acceleration in the remote past and, respectively, future,
 we can again paste
 together the corresponding retarded and advanced isolated color source
 solutions of \cite{mikhailov,noline} to obtain a valid quark-antiquark embedding.

 For the embedding to be smooth at the juncture between the retarded and advanced solutions (\ref{mikhsolnoncovariant}), we need to demand that the asymptotic past acceleration of the quark, $A_q$, matches the asymptotic future acceleration of the antiquark, $A_{\bar{q}}$. Physically valid quark-antiquark embeddings are also obtained for $A_q\neq A_{\bar{q}}$, but they are not smooth: the radii of the corresponding asymptotic circles (\ref{xiaoaccel}) do not match, and the region in between needs to be completed with a vertical string extending from $z=A^{-1}_q$ to $z=A^{-1}_{\bar{q}}$ and always moving at the speed of light. As explained in \cite{noline}, this vertical segment encodes a portion of a gluonic shock wave, which in the present case, is shed from the quark and ultimately absorbed by the antiquark. The non-smooth character of the solution for $A_q\neq A_{\bar{q}}$ is entirely due to the fact that the endpoint velocities asymptotically approach the speed of light. If instead one considers the more physical situation where the asymptotic speeds are $1-\epsilon$, with any $0<\epsilon\ll 1$, then the embeddings are smooth \cite{noline}. As $\epsilon\to 0$, such embeddings continuously revert to the non-smooth case that we have contemplated in this paragraph.

This vastly enlarged class of analytic solutions contains in particular all cases where \emph{both} the quark and antiquark asymptote to uniform acceleration in the past \emph{and} future, but have arbitrary worldlines in between. This is the full collection of trajectories envisioned by Jensen and Karch, where the quark and antiquark are restricted to lie within complementary Rindler wedges, and are therefore permanently out of causal contact in spacetime. As anticipated in \cite{jensenkarch}, the corresponding worldsheets inherit this same structure: they contain a double-sided horizon (of white/black hole character for $t<0$ and $t>0$, respectively) that forever precludes propagation of signals along the string between the two endpoints.  An example is
 shown in Fig.~\ref{accel1fig}. The expressions needed to trace the horizon can be found in the Appendix. 


  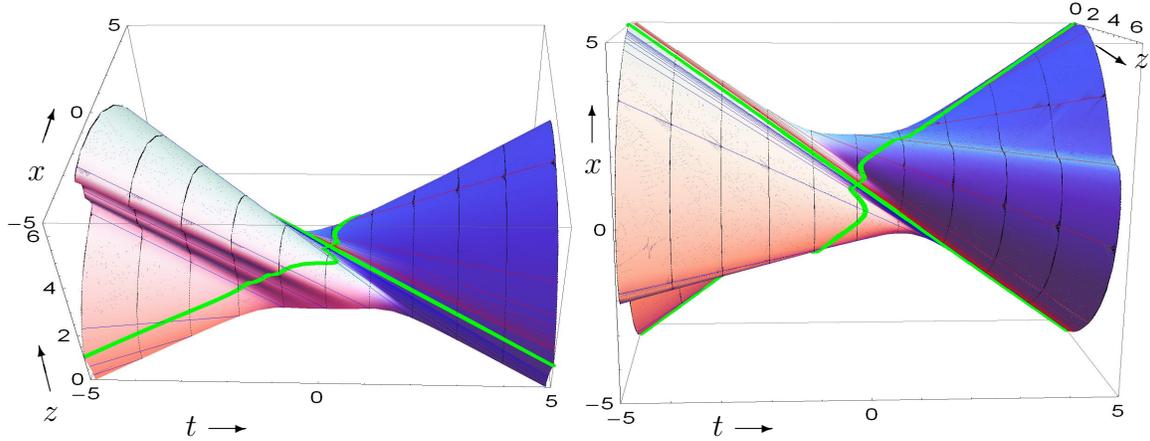
\begin{figure}[tbph]
\begin{center}
\vspace*{0.5cm}
 \setlength{\unitlength}{1cm}
\includegraphics[width=7.5cm,height=5.7cm]{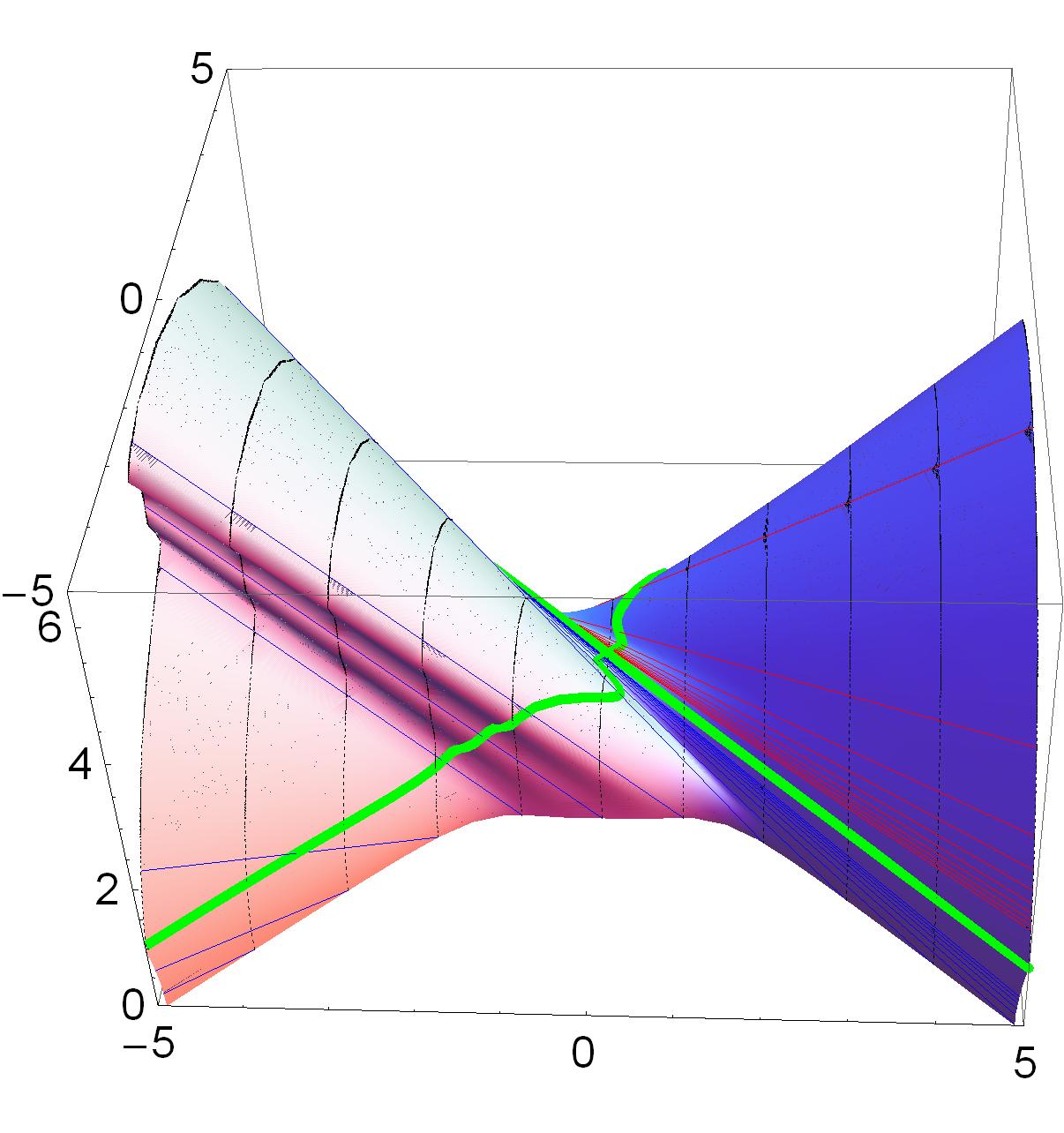}
\includegraphics[width=7.5cm,height=5.7cm]{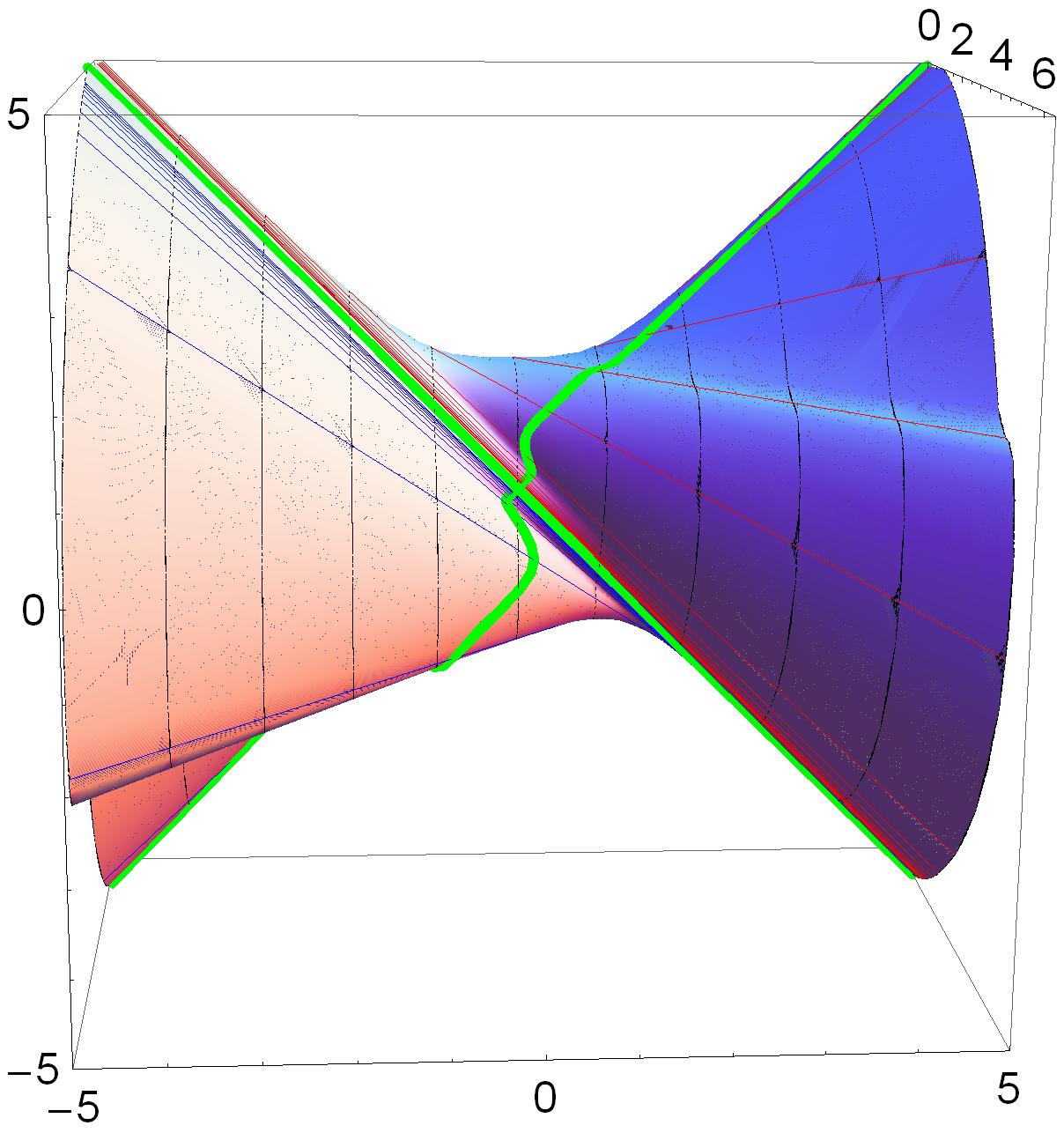}
 \begin{picture}(0,0)
   \put(-7.3,3.8){$x$}
   \put(-7.1,4.3){\vector(1,3){0.15}}
   \put(-7.1,0.6){$z$}
   \put(-7.0,1.0){\vector(-1,4){0.15}}
   \put(-5.2,0.4){$t$}
   \put(-4.9,0.5){\vector(1,0){0.5}}
   \put(0.1,3.9){$x$}
   \put(0.2,4.3){\vector(0,1){0.5}}
   \put(7.4,5.3){$z$}
   \put(6.9,5.6){\vector(3,-2){0.4}}
   \put(1.8,0.4){$t$}
   \put(2.1,0.5){\vector(1,0){0.5}}
  \end{picture}
 \end{center}
 \vspace*{-0.5cm}
\caption{Side and top views of the analytic embedding for a string dual to an infinitely massive quark and antiquark which asymptote to hyperbolic motion both in the past and in the future, with the same uniform acceleration $A$. Positions in the graphs are given in units of $A^{-1}$. Dotted contours are snapshots of the string at fixed time. The diagonal straight line at $z=A^{-1}$, running from the top left to the bottom right of the figures (green online), is one side of the worldsheet horizon. The region to its right (mostly dark) is the retarded embedding (\ref{mikhsolnoncovariant})  for quark trajectory $x_q(t)=\sqrt{A^{-2} + t^2 + \exp(-A^2 t^2)}$. The region to the left (mostly light) is the advanced solution (\ref{mikhsolnoncovariant}) for antiquark trajectory $x_{\bar{q}}(t)=-\sqrt{A^{-2} + t^2} -0.9 \exp(-A^2 t^2) \sqrt{A^{-2}\cos(-A^{2}t^2/2) + t^2}$. The outgoing/ingoing null geodesics which form the advanced/retarded worldsheet (\ref{mikhsolnoncovariant}) are seen to be straight lines (blue/red online), and to cluster around the mentioned horizon. The other side of the horizon is the somewhat tortuous curve running across the opposite diagonal (also green online). It lies at height $z=A^{-1}$ in the remote past and future, as well as in the immediate vicinity of $t=0$, but deviates from that in the region associated with non-uniform acceleration. The endpoints are seen to be causally disconnected along the string at all times, just as they are along spacetime.}\label{accel1fig}
\end{figure}

  \begin{figure}[tbph]
\begin{center}
\vspace*{0.5cm}
 \setlength{\unitlength}{1cm}
\includegraphics[width=7.5cm,height=5.7cm]{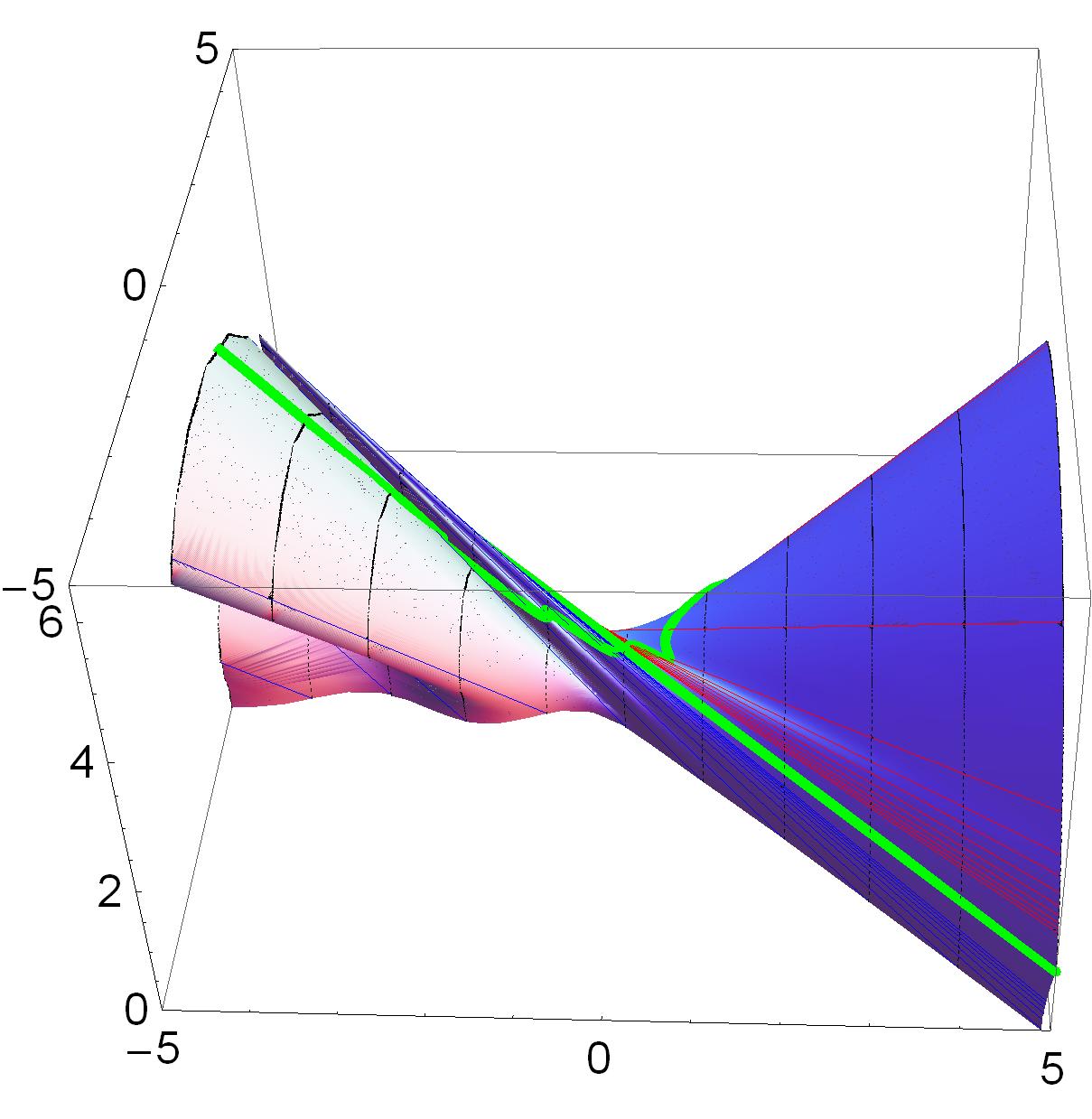}
\includegraphics[width=7.5cm,height=5.7cm]{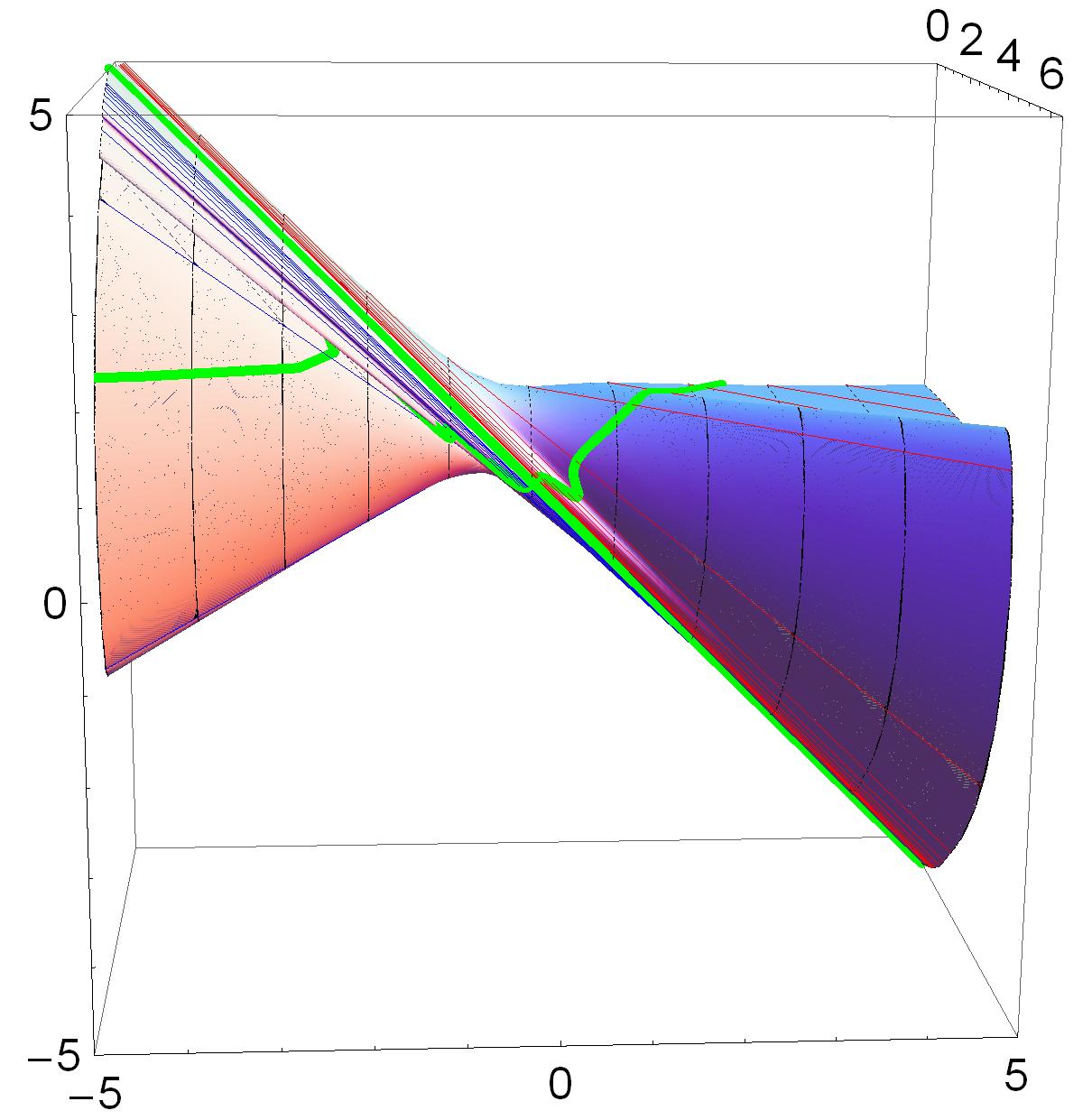}
 \begin{picture}(0,0)
   \put(-7.3,3.8){$x$}
   \put(-7.1,4.3){\vector(1,3){0.15}}
   \put(-7.1,0.6){$z$}
   \put(-7.0,1.0){\vector(-1,4){0.15}}
   \put(-5.2,0.4){$t$}
   \put(-4.9,0.5){\vector(1,0){0.5}}
   \put(0.1,3.9){$x$}
   \put(0.2,4.3){\vector(0,1){0.5}}
   \put(7.4,5.3){$z$}
   \put(6.9,5.6){\vector(3,-2){0.4}}
   \put(1.8,0.4){$t$}
   \put(2.1,0.5){\vector(1,0){0.5}}
  \end{picture}
 \end{center}
 \vspace*{-0.5cm}
\caption{Side and top views of the analytic embedding for a string dual to an infinitely massive quark (and antiquark) which asymptotes to hyperbolic motion with acceleration $A$ in the past (future), but is static in the future (past). Positions in the graphs are given in units of $A^{-1}$. Dotted contours are snapshots of the string at fixed time. The diagonal straight line at $z=A^{-1}$, running from the top left to the bottom right of the figures (green online), is one side of the worldsheet horizon. The region to its right (mostly dark) is the retarded embedding (\ref{mikhsolnoncovariant})  for quark trajectory $x_q(t)= A^{-1}[1+(1/2)(1 - \tanh t ) (\sqrt{A^{-2} + t^2} - 1)]$. The region to the left (mostly light) is the advanced solution (\ref{mikhsolnoncovariant}) for antiquark trajectory $x_{\bar{q}}(t)=
A^{-1}[(1/3) \exp(-A^2 t^2/25) \cos(2 (t - b)) +
 (1/2)(1 + \tanh t) (\sqrt{A^{-2} + t^2} -
    (1/3) \exp(-A^2 t^2/25) \cos(2 (t - b))]$, with $b=1.3A^{-1}$. The outgoing/ingoing null geodesics which form the advanced/retarded worldsheet (\ref{mikhsolnoncovariant}) are seen to be straight lines (blue/red online), and to cluster around the mentioned horizon. The other side of the horizon is the tortuous curve running across the opposite diagonal (also green online). It descends from $z\to\infty$ in the remote past, crosses $z=A^{-1}$ at $t=0$, and reaches the string endpoint corresponding to the quark at $t=2.434 A^{-1}$, $x=1.012 A^{-1}$.
     The endpoints are seen to be causally disconnected along the string at all times, even though along spacetime the antiquark can send signals to the quark.}\label{accel2fig}
\end{figure}

The other embeddings we have constructed here are also interesting, because they include cases where one or both particles are not confined to their corresponding Rindler wedge. If the antiquark is not uniformly accelerated in the remote past, and/or the quark is not uniformly accelerated in the far future, then in spacetime signals can travel from the antiquark to the quark, and not the opposite. But the worldsheet is still found to contain a double-sided horizon of white and black hole character for $t<0$ and $t>0$, respectively, meaning that the endpoints are forever causally disconnected along the string, in either direction. An example is given in Fig.~\ref{accel2fig}.
In these cases, then, the situation from the spacetime and worldsheet perspectives is \emph{not} the same.

We can also consider a quark and antiquark that asymptote to uniform acceleration in the \emph{same} Rindler wedge (an example that was briefly mentioned in \cite{jensenkarch}). In this case we again have $A_q\neq A_{\bar{q}}$, so as above, the region in between the two horizons must be filled out by a vertical string segment \cite{noline}. The mismatch between the two perspectives on causality  is more poignant in these examples: even though the $q$ and $\bar{q}$ are never out of bidirectional contact in spacetime, they are permanently disconnected along the string.

It is worth emphasizing that, given a quark that asymptotes to hyperbolic motion (in either the past or the future), we are certainly not \emph{required} to complete the dual string worldsheet in such
a way that it returns to the flavor branes at $z=z_m$, implying the existence of an antiquark. As explained in \cite{noline}, the relevant solutions (\ref{mikhsolnoncovariant}) allow a second type of continuation, appropriate for an isolated color charge. In these cases, a portion of the semicircular embedding (\ref{xiaoaccel}) is attached to a vertical string extending all the way to the Poincar\'e horizon, and moving at the speed of light. As mentioned above, from the MYSM perspective this segment represents a gluonic shock wave that is shed (absorbed) by the quark (antiquark). The corresponding worldsheets will then evidently have a single region exterior to the horizon at $z=A^{-1}$.


Notice that in all cases covered in this section, including the example of strictly uniform acceleration \cite{xiao}, there is an interesting novelty with respect to the more physical examples of Section \ref{freesec}. For pair histories where both in the past and in the future we have at least one particle undergoing hyperbolic motion, the region enclosed by the worldsheet horizon, even at $t\to\infty$, is found \emph{not} to be associated exclusively with the emitted radiation. For instance, in the embedding (\ref{xiaoaccel}), depicted in Fig.~\ref{xiaofig}, the horizon is always at $z=A^{-1}$, so the exterior regions at $t\to\infty$ do not match the known dual to the asymptotic quark/antiquark: they are indeed vertical and moving at the speed of light, but they are missing the portion running from $z=A^{-1}$ to $z\to\infty$, which in MSYM language represents the IR part of the expected gluonic shock wave. Similarly, in the example portrayed in Fig.~\ref{accel2fig}, for $t>2.434 A^{-1}$ the worldsheet horizon engulfs the entire side of the string dual to the quark, even though this particle simply asymptotes to a resting configuration.

If we want, we can certainly delineate the worldsheet region that contains the radiation at $t\to\infty$ by focusing on the null geodesics that mark the boundary of the complete portions dual to the asymptotic quark and antiquark, which \emph{are} correctly present in all embeddings. (Again, the relevant formulas can be found in the Appendix.) Such an analysis would define an inner horizon that, in the class of solutions constructed in this section, does not agree with the outer horizon that is relevant for signals originating from the string endpoints. This mismatch is due to the the violent nature of the initial and final conditions associated with uniform acceleration, which could only arise in the presence of eternal forcing. We thus understand why there was no such mismatch for the more physical situations described in Section \ref{freesec}. Notice in particular that in the hyperbolic examples the quark and/or antiquark never cease to radiate, so there is no time where it becomes easy to disentangle the `near' and `far' contributions to the gluonic field.

\section{Other Examples} \label{othersec}

In the previous section we discovered a large class of examples where analytic quark-antiquark solutions (including the one \cite{xiao} used in \cite{jensenkarch})
can be constructed directly by joining together the purely retarded/advanced solutions (\ref{mikhsolnoncovariant}) found in \cite{mikhailov}, which, separately, would describe an isolated quark or antiquark. The key ingredient that allowed this was the presence of a worldsheet horizon,
which prevents causal contact between the two string endpoints. It stands to reason, then, that in cases where radiation is emitted, and consequently a double-sided horizon is formed on the worldsheet, the exterior regions of the string, associated with the quark and antiquark, should asymptotically approach the isolated solutions (\ref{mikhsolnoncovariant}). We in fact see this explicitly in the cases discussed in Section \ref{freesec}, where at $t\to\infty$ the string develops two infinite vertical segments translating uniformly, which are described precisely
by the instance of (\ref{mikhsolnoncovariant})
with constant endpoint velocity.

This means that, whenever the quark-antiquark system radiates, we know the late-time form of the two regions exterior to the worldsheet horizon. Consider for instance a pair that is static (thanks to an appropriate external force) up to some time $t=t_{\mbox{\scriptsize pull}}$, after which the quark and antiquark are pulled away from one another up to $t=t_{\mbox{\scriptsize hold}}$, and are subsequently held fixed again. If enough energy is injected in the intermediate stage, the quark and antiquark will become unbound, and radiation will be emitted. This implies that, in the gravity description, a double-sided horizon will form on the worldsheet, and the string will ultimately evolve to two static vertical strings (dual to the color sources) connected by a horizontal segment (dual to the radiation) arbitrarily close to $z\to\infty$.
Notice that, here and in all cases where a black hole is formed, the fact that we are working in the $\lambda\to\infty$ limit means that it cannot evaporate---worldsheet Hawking radiation\footnote{The effects of such Hawking radiation were first discussed in the finite temperature context \cite{gubserqhat,ctqhat,deboerhubeny,sonteaney}.} would be a $1/\sqrt{\lambda}$ effect \cite{xiao,brownian,hkkl}.

Another example of the same type would be to take an initially static pair of length $L$ and rotate it around its midpoint, with uniform speed $v$, forever after. This requires an infinite amount of energy to be supplied to the system, which unbinds the pair and flows away as radiation. The asymptotic string configuration will then be a horizontal highly coiled radiation region at $z\to\infty$, connecting  two copies of the embedding (\ref{mikhsolnoncovariant}) for uniform circular motion. The latter in fact agrees with the spiral string independently constructed in \cite{liusynchrotron} to study synchrotron radiation. In this case the worldsheet horizon is known to be at the fixed radial position
$z=L/2\gamma^2 v^2$, so again, the eternal forcing produces a situation where the regions exterior to the outer horizon do not match the full profiles expected for an isolated quark/antiquark. Analogous conclusions can be drawn for situations where the quark and antiquark are made to oscillate periodically (either jointly or separately) in various ways (e.g., the version of (\ref{mikhsolnoncovariant}) for harmonic oscillation was studied in \cite{trfsq}).

An important property of all these examples is that, just as in the last category of embeddings analyzed in the previous section (such as the one in Fig.~\ref{accel2fig}), the string endpoints remain in causal contact through spacetime even after they become disconnected along the string.

One final example we would like to mention is the following. With the intention of modeling pair creation in the midst of a thermal plasma, the work \cite{hkkky} studied the evolution of a quark and antiquark that begin at the same position at $t=0$ and then (without external forcing) separate back to back. The dual embeddings considered there described an initially pointlike string, which became $\cap$-shaped and progressively descended into the bulk of AdS as its endpoints separated. The work \cite{dragtime} later pointed out that a second way to satisfy the initial conditions and constraint equations for the string is to have a $\wedge
$-shaped embedding with its vertex at the spacetime horizon, which starts out as a (folded up) line at $t=0$ and begins to open up as its endpoints separate. It was argued that, whereas the initially pointlike string is dual to the $q$-$\bar{q}$ pair in a color-singlet configuration (which could have been created, e.g., by a photon), the string that is initially linelike corresponds to the adjoint configuration of the quark and antiquark (which could have been created by a gluon). At leading order in $1/N_c$, the latter is unbound and indistinguishable from an isolated quark and antiquark which just happen to be
initially at the same point, but the constraint analysis in \cite{dragtime} led to a particular set of velocity profiles for the string, which translate into particular initial gluonic profiles. To the extent that one is willing to interpret (the $T=0$ version of) this configuration as an adjoint quark-antiquark pair, it provides another example where the two color sources are entangled and are always in causal contact through spacetime, even though they are permanently disconnected along the string. (Since the two halves of the string evolve independently, they are particular instances of (\ref{mikhsolnoncovariant}), and their worldsheet horizons can therefore be traced as in \cite{dragtime} and the examples of the previous section.) It is perhaps worth emphasizing that the permanent disconnection on the worldsheet is not just a feature of these adjoint solutions, but also of those where both particles have asymptotically uniform acceleration, such as (\ref{xiaoaccel}).

\section{Global AdS} \label{globalsec}

It was shown in \cite{jensenkarch} that, upon changing from Poincar\'e to global coordinates in AdS, the expanding circular
string (\ref{xiaoaccel}) dual to a uniformly accelerating $q$-$\bar{q}$ pair on $\mathrm{R}^{3,1}$ becomes a static string stretching straight across AdS, dual to a pair on  MSYM on $S^3\times \mathrm{R}$, with the $q$ and $\bar{q}$ at rest at antipodal points of the $S^3$. The definition of the region to which signals on the string can escape changes drastically under this mapping, so the causal structure of the worldsheet (as well as of spacetime itself) is different when interpreted from the Poincar\'e and global perspectives. In particular, for the static pair in global AdS, there are no worldsheet horizons.

\begin{figure}[htb]
\setlength{\unitlength}{1cm}
\begin{center}
\includegraphics[width=5cm]{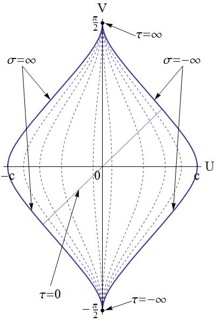}
\hspace*{2cm}
\includegraphics[width=5cm]{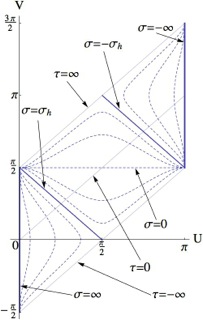}
\end{center}
	\caption{Penrose diagrams (constructed in \cite{mariano}) of the worldsheet
	for the string dual to a quark-antiquark pair in global AdS rotating on a non-maximal circle with low (left) or high (right) angular velocity. The two endpoints of the worldsheet correspond to the quark (at $\sigma=\infty$) and the antiquark
	(at $\sigma=-\infty$), which we depict with thick lines. In the first plot, they intersect
	the horizontal axis at $U=\pm c=\pm \arctan (\pi/2)$. The lines at $\sigma=\pm\sigma_h$ in the second plot are worldsheet horizons.
Dashed curves are other lines with constant
	$\sigma$.
	Null trajectories going to the right correspond to constant $\tau$ in both plots.
The main feature is that, when the causal structure is non-trivial, signals can flow from the quark to the antiquark, but not the other way around. See \cite{mariano} for details.}
	\label{fig:penrose}
\end{figure}

Beyond this particular example of a static pair, an infinite family of other $q$-$\bar{q}$ trajectories on $S^3\times \mathrm{R}$ (including the global translation of all the cases analyzed in Section \ref{accelsec}) can be accessed through  \cite{mikhailov}, where the retarded/advanced solutions (\ref{mikhsolnoncovariant}) were in fact constructed first in global coordinates. The worldsheet causal structure for such embeddings was studied
in \cite{mariano}, which (in the interest of exploring the analog of the Unruh effect for circular motion) focused primarily on the class of solutions where the quark rotates uniformly along a circular path at arbitrary latitude on the $S^3$, and the antiquark does the same but along the antipodal circle. It was found that a double-sided worldsheet horizon appears for sufficiently large angular velocity ($\omega>1/\sqrt{bR_0}$, with $b$ and $R_0$ respectively the $S^3$ and orbit radius). The nature of the horizons is such that signals can propagate only from (say) the quark to the antiquark, but not the reverse (see the left Penrose diagram in Fig.~\ref{fig:penrose}), so unlike what happens in the Poincar\'e solutions discussed in previous sections, here the region between the two horizons is not trapped.
In the MSYM language, this is related to the fact that, on a sphere, there is no unambiguous notion of radiation, and what one sees is a continuous flow of energy from the quark (where it is injected by an external agent) to the antiquark (where it is extracted).

\section{Discussion}\label{discussionsec}

 In this paper we have considered the string that is holographically dual to an entangled quark-antiquark pair, and studied its causal structure for many different types of trajectories. Included were the cases with asymptotically uniform acceleration (Section \ref{accelsec}), where based on \cite{mikhailov,noline} we have constructed an infinite class of analytic solutions generalizing the one obtained in \cite{xiao}. Additionally, we have analyzed the cases where the pair is asymptotically free (Section \ref{freesec})
or forced in some steady manner (Section \ref{othersec}), for which we do not have the explicit embeddings but can nevertheless
 infer the worldsheet structure from knowledge of the asymptotic behavior of the pair. We also recalled in Section \ref{globalsec} the results of \cite{mariano} for string embeddings in global AdS.

 {}From our worldsheet analysis we have extracted three main conclusions. First, the existence of a double-sided horizon with what one could call a wormhole in between is not specific to the uniformly accelerated case examined in \cite{jensenkarch} (or the larger class of asymptotically uniformly accelerated worldlines envisioned there), but is a feature of rather generic trajectories in Poincar\'e AdS. Importantly for the ER=EPR story, all of these wormholes are non-traversable, because the two horizons they connect are outgoing null geodesics, and so, by definition, can only be crossed by ingoing fluctuations. As we saw, the basic physics involved in their formation is really the injection of energy into the string, which then collapses to form a black hole on the worldsheet. The fact that there are two exterior regions (corresponding to the 2 string endpoints) instead of one as in a standard black hole formed by collapse is due to one-dimensional character of the space: a black hole in the middle of a line necessarily leads to causal disconnection of the segments on either side. The situation in the example of \cite{jensenkarch} is fundamentally no different.
 
 Second, and related to the previous point, the question of whether or not the string endpoints are causally connected along the body of the string is distinct from whether they are in causal contact in the spacetime sense. We have seen many examples where the endpoints cannot communicate via the string even though they remain within reach of one another through spacetime. In MSYM language, the distinction is that, whereas the spacetime metric controls the propagation of generic signals, the information in the worldsheet metric determines whether or not signals can be sent between the quark and antiquark \emph{in the form of perturbations of the gluonic field}. In other words, the presence of the excited gluonic medium translates into a spatiotemporally varying refraction index that modifies the causal structure seen
 by gluonic waves.
 
 Third, the crucial ingredient needed for a non-trivial worldsheet causal structure (at $N_c,\lambda\to\infty$) is the emission of gluonic radiation: whenever the $q$-$\bar{q}$ pair radiates, it becomes unbound and a double-sided horizon develops on the dual string.
 In the distant future, and for the more physical examples where forces are not applied eternally, this horizon demarcates an interior region of the string, which encodes the radiation, from the two exterior regions, which correspond to the quark and antiquark. The intimate connection between gluonic radiation and worldsheet horizons was first discovered in \cite{dragtime} for the case of isolated color sources; here we have learned that it holds also for quark-antiquark pairs. For situations with eternal forcing, such as the asymptotically uniformly accelerated pair histories studied in Section \ref{accelsec}, we saw that the region ultimately enclosed by the outermost worldsheet horizon again encodes radiation, but may also contain a portion of the gluonic `near' field.\footnote{It is also worth noting that, even after forcing is turned off, it is not possible \emph{at finite time} to literally identify  the interior of the horizon with the radiation and the exterior with the quark and antiquark \cite{noline}. The horizon ends up playing the role of dividing line between the color sources and the emitted radiation (in the asymptotically free cases) only for $t\to\infty$. Still, what is clear is that the appearance of a horizon signals the emission of radiation, and the region enclosed by the horizon can thus be thought of as `proto-radiation'.}

 Before we attempt to draw lessons from our findings regarding the possible relation with the ER=EPR conjecture,
it is perhaps useful to note that the perspective of \cite{jensenkarch} seems to be somewhat different from that of \cite{maldasuss} in two respects. Whereas Maldacena and Susskind mostly argue for a connection between wormholes and entanglement in one and the same quantum gravity theory, Jensen and Karch highlight a situation where the AdS/CFT correspondence appears to be translating entanglement in a non-gravitational theory into a (worldsheet) wormhole in the gravity dual. Additionally, the authors of \cite{jensenkarch} impose the restriction that the members of the EPR pair are permanently out of causal contact, whereas no such condition is apparent to us in the widest-ranging (and most speculative) version  of the conjecture of \cite{maldasuss}.

 Now, what implications does the horizon-radiation connection established in this work have for the interpretation of holographic EPR pairs put forth in \cite{jensenkarch}? Our results support the view that the existence of a worldsheet horizon is tied to entanglement, in two different but related senses. On the one hand, the quantum fluctuations of the string embedding fields $\delta \vec{X}(t,z)$ on either side of the horizon are entangled, which (at order $1/\sqrt{\lambda}$) leads in particular to worldsheet Hawking radiation. This in turn gives rise to stochastic motion of the endpoints, just as is expected to result in MSYM from the process of gluonic radiation, or, alternatively, in the rest frame of the quark, from the Unruh-type thermal medium \cite{xiao,brownian}. On the other hand, the fact that as $t\to\infty$ the horizon delineates the region of the string that codifies the emitted radiation implies that, from the MSYM perspective, the existence of the horizon is indicative of entanglement between the particles and the radiation.

 Indeed, the main lesson underscored by our findings is that the system under study is much more complex than a standard EPR \emph{pair}. This had been emphasized already in \cite{jensenkarch}, in the sense that the quark and antiquark should be understood as dressed quasiparticles, each including its corresponding gluonic cloud, or in other words, the `near' portion of the gluonic field it generates \cite{martinfsq,lorentzdirac,damping}. Here we have learned that  the complexity of the system is greatly compounded by the fact that the string additionally encodes the `far' gluonic field, which by its very nature is much more cleanly disjoint from the (dressed) quark and antiquark. Concretely, then, the issue is that, having begun with a color-singlet $q$-$\bar{q}$ pair, by the time we unbind the members of the pair to be able to consider them as separate entities, part of the color has been radiated away, so the relevant correlations hold not between just the quark and the antiquark, but between the quark and the joint radiation $+$ antiquark system. It is this entanglement that is codified by the presence of the worldsheet horizon.\footnote{A curious feature is that, whereas the wormholes discussed in \cite{maldasuss} are the innermost portion of the spacetime geometry, inaccessible to exterior observers, the wormholes that appear on the string worldsheet always codify radiation, and therefore end up being the outermost component of the system in the MSYM description, readily accessible to distant observers.}

 The situation is roughly analogous to an electron and positron that start out as a standard EPR pair in the sense that their spins are entangled, but are then separated while being allowed to radiate electromagnetically. In the end, knowing the spin of the positron would clearly not suffice to know the spin of the electron, because angular momentum will in general have been carried away by the emitted photons. Viewed in this light, it is actually quite remarkable that the AdS/CFT framework allows the correlations of the highly complicated strongly-coupled non-Abelian `pair' to be
  reexpressed in terms of the causal structure of the dual string. And we also understand why the detailed form of the entanglement (encoded in the detailed geometry of the wormhole region) depends on the specific choice of trajectories for the quark and antiquark.

 The fact that the class of $q$-$\bar{q}$ trajectories where there is no horizon are precisely those in which the system remains bound can also be given an interpretation along these same lines: in those cases, there is no inherent partition of the pair into components, so it seems natural that no intrinsic subdivision is found to appear in the worldsheet geometry.
 It is only with the emission of radiation and the consequent unbinding of the pair that one obtains an unambiguous separation into subsystems that are entangled, and this is what is reflected on the gravity side by the existence of the worldsheet horizon.

For the cases where the quark and antiquark asymptotically undergo uniform acceleration (for which we constructed analytical embeddings in Section \ref{accelsec}), the authors of \cite{jensenkarch} were able to trace a link between the worldsheet horizon and the entanglement entropy of the system. Their argument involved a change of coordinates in AdS, to a hyperbolic slicing where the spacetime Rindler horizon is reinterpreted as a thermal horizon, and the single endpoint of the string that remains visible is found to be at rest \cite{xiao,brownian}. In this setting the contribution of the string to the entanglement entropy \cite{rt} can be computed as a thermal entropy \cite{chm}, using the logic of \cite{kot}. It would be interesting if the entanglement entropy could be determined for the much more generic cases which we have discussed in this paper, where the horizon is highly dynamical and thermal entropy is not likely to be a useful concept. In all cases, the correlations due to entanglement are reflected
in non-vanishing connected correlators for fluctuations of the two endpoints, as discussed in \cite{jensenkarch}. An interesting novelty is that, based on our results, there will be situations where the \emph{causal} quark-antiquark correlators vanish due to the worldsheet horizons, even though the locations of the two particles are within reach of one another through spacetime. This is a concrete indicator of the fact, already emphasized above, that two different notions of causal structure are at play here.

\section*{Acknowledgements}
We are grateful to Kristan Jensen and Andreas Karch for a useful email exchange on \cite{jensenkarch}, which got us started on this project. We also thank our JHEP referees for constructive suggestions on the text. MC is supported by the European Research Council under the European Union's Seventh Framework Programme (FP7/2007-2013), ERC grant agreement STG 279943, ``Strongly Coupled Systems''.
AG is partially supported by Mexico's National Council of Science and Technology (CONACyT) grant 104649, as well as DGAPA-UNAM grant IN110312.
JFP is partially supported by the National Science Foundation under Grant No. PHY-0969020 and by the Texas Cosmology Center.

 \section*{Appendix}

In the static gauge $\tau=t$, $\sigma=z$, the induced metric on the string reads
\begin{equation}\label{indmetric}
ds^2=\frac{R^2}{z^2}\left[\left(-1+\dot{\vec{X}}^2\right)dt^2
+2 \dot{\vec{X}}\cdot\vec{X}^{\prime}
+\left(1+\vec{X}^{\prime 2}\right)dz^2\right]~.
\end{equation}
We see immediately from this that, if there exists a locus $z_{\mbox{\scriptsize
s}}(t)$ where $\dot{\vec{X}}^2=1$, then it is a stationary-limit curve ($g_{tt}=0$): by causality, signals at such points along the string cannot move outwards (towards smaller $z$). More generally, (\ref{indmetric}) implies that the ingoing and outgoing null geodesics on the string respectively obey the differential equation
\begin{equation}\label{znulldot}
\dot{z}_{\mbox{\scriptsize
null}}^{(\pm)}(t)=\frac{-\dot{\vec{X}}\cdot\vec{X}^{\prime}
\pm\sqrt{1-\dot{\vec{X}}^2+\vec{X}^{\prime 2}
+\left(\dot{\vec{X}}\cdot\vec{X}^{\prime}\right)^2-\dot{\vec{X}}^2\vec{X}^{\prime 2}}}{1+\vec{X}^{\prime 2}}~.
\end{equation}
It is easy to check here that indeed $\dot{\vec{X}}^2=1$ translates into  $\dot{z}_{\mbox{\scriptsize
null}}^{(-)}(t)=0$, meaning that the lightcone that would normally point outward is now horizontal, which is the condition that identifies the stationary-limit curve.

In the cases discussed in Section \ref{accelsec}, where the string embedding is constructed by pasting together a retarded and an advanced copy of (\ref{mikhsolnoncovariant}) for one-dimensional motion, the equations determining the null geodesics on the worldsheet can be worked out explicitly in terms of the quark and antiquark data. The ingoing (outgoing) null curves on the retarded (advanced) portions are directly given by the straight lines at constant $t_r$, i.e., $\dot{z}^{(\pm)}_{\mbox{\scriptsize
null}}(t)
=\pm\sqrt{1-v^2}$. The remaining, nontrivial null geodesics are the outgoing ones on the retarded portion and ingoing on the advanced piece, which respectively obey
\begin{equation}\label{znullmikh}
\dot{z}_{\mbox{\scriptsize null}}^{(\mp)}(t)
=\frac{\mp\sqrt{1-v^2}\left((1-v^2)^3-a^2 z^2\right)}{(1-v^2)^3+a^2 z^2
\pm 2 vaz(1-v^2)^{3/2}}~.
\end{equation}
Using (\ref{mikhsolnoncovariant}) again, this can be seen to translate into a rate of change with respect to $t_r$ equal to
$dz_{\mbox{\scriptsize null}}^{(\mp)}/dt_r=\mp((1-v^2)^3-a^2 z^2)/2(1-v^2)^{3/2}$. {}From these expressions we learn in particular that $z^2=(1-v^2)^3/a^2$ marks the location of the stationary-limit curve.

We now return to the general discussion. The metric (\ref{indmetric}) controls the propagation of small perturbations on the given worldsheet, and also the flow of spacetime momentum along the string.  The momentum densities following from (\ref{nambugoto}) read
\begin{eqnarray}\label{momenta}
\Pi^t_t&=&-\frac{\sqrt{\lambda}}{2\pi}
\frac{1+\vec{X}^{\prime 2}}{z^2\sqrt{1-\dot{\vec{X}}^2+\vec{X}^{\prime 2}
+\left(\dot{\vec{X}}\cdot\vec{X}^{\prime}\right)^2-\dot{\vec{X}}^2\vec{X}^{\prime 2}}}~,\nonumber\\
\Pi^t_{\vec{x}}&=&\frac{\sqrt{\lambda}}{2\pi}
\frac{\left(1+\vec{X}^{\prime 2}\right)\dot{\vec{X}}
-\left(\dot{\vec{X}}\cdot\vec{X}^{\prime}\right)\vec{X}^{\prime}}{z^2 \sqrt{1-\dot{\vec{X}}^2+\vec{X}^{\prime 2}
+\left(\dot{\vec{X}}\cdot\vec{X}^{\prime}\right)^2-\dot{\vec{X}}^2\vec{X}^{\prime 2}}}~,\\
\Pi^z_t&=&\frac{\sqrt{\lambda}}{2\pi}
\frac{\dot{\vec{X}}\cdot\vec{X}^{\prime}}{z^2\sqrt{1-\dot{\vec{X}}^2+\vec{X}^{\prime 2}
+\left(\dot{\vec{X}}\cdot\vec{X}^{\prime}\right)^2-\dot{\vec{X}}^2\vec{X}^{\prime 2}}}~,\nonumber\\
\Pi^z_{\vec{x}}&=&-\frac{\sqrt{\lambda}}{2\pi}
\frac{\left(1-\dot{\vec{X}}^2\right)\vec{X}^{\prime}
+\left(\dot{\vec{X}}\cdot\vec{X}^{\prime}\right)\dot{\vec{X}}}{z^2\sqrt{1-\dot{\vec{X}}^2+\vec{X}^{\prime 2}
+\left(\dot{\vec{X}}\cdot\vec{X}^{\prime}\right)^2-\dot{\vec{X}}^2\vec{X}^{\prime 2}}}~.\nonumber
\end{eqnarray}
It can be deduced from this and (\ref{znulldot}) that the flow of energy across any outward null geodesic, $dE/dt\equiv -(\Pi^z_t-\dot{z}\Pi^t_t)$, is as expected positive definite, and equal to $\sqrt{\lambda}/2\pi z^2$.

The worldsheet horizon $z_{\mbox{\scriptsize
h}}(t)$, when it exists, is the particular outward geodesic $z_{\mbox{\scriptsize
null}}^{(-)}(t)$ that marks the boundary of the escape region. In the case of asymptotically free motions discussed in Section \ref{freesec}, the escape region is given by the infinite vertical string segments dual to the uniformly translating quark and antiquark, so $z_{\mbox{\scriptsize
h}}(t)$ must retreat to $z\to\infty$ as $t\to\infty$. While $z_{\mbox{\scriptsize
h}}(t)$ is moving inward in this way, it must necessarily lie above (the outer portion of) the stationary-limit curve
$z_{\mbox{\scriptsize
s}}(t)$. The latter is of course easier to find than the horizon, because it is defined by a local condition on the worldsheet.

If at $t=t_{\mbox{\scriptsize pull}}$ we start with a configuration that does not have a horizon, then by pulling on the endpoints we will induce some interior point on the string to reach the speed of light ($|\dot{\vec{X}}|=1$). {}From this moment on, there will be a stationary-limit curve $z_{\mbox{\scriptsize
s}}(t)$ marking the boundary of the superluminal region of the string.\footnote{Of course, the physical, \emph{transverse} velocity of the string never becomes superluminal.} After the endpoints are released at $t=t_{\mbox{\scriptsize free}}$, this superluminal region will contract and move towards $z\to\infty$. The horizon $z_{\mbox{\scriptsize
h}}(t)$ must be found within it, and can be mapped out, for any given $q$-$\bar{q}$ trajectory, by starting at $t\to\infty$ and following the corresponding outward null geodesic backwards. The situation is essentially a double-sided version of the one discussed in Section 2.3 of \cite{dragtime} (see Fig.~1 of that paper).

If instead of releasing the endpoints we continue forcing them appropriately up to $t\to\infty$, then we can produce situations where neither the stationary-limit curve nor the horizon retreats to $z\to\infty$. Such cases are contemplated in Sections \ref{accelsec} and \ref{othersec}. In particular, when both the quark and antiquark are asymptotically uniformly accelerated, or permanently forced in some periodic trajectory such as uniform circular motion, one finds that
$z_{\mbox{\scriptsize s}}(t)=\mbox{constant}$, and therefore $z_{\mbox{\scriptsize h}}(t)=z_{\mbox{\scriptsize s}}(t)$.


\end{document}